\documentclass[twocolumn,tighten]{aastex62}
\usepackage{wasysym}

\newcommand{\sion}[2]{#1$\;$\textsc{#2}\relax}
\newcommand{\sub}[2]{\ifmmode #1_\mathrm{\scriptstyle #2} \else $#1_\mathrm{\scriptstyle #2}$\fi}
\newcommand{\ssub}[2]{\ifmmode #1_\mathrm{\scriptscriptstyle #2} \else $#1_\mathrm{\scriptscriptstyle #2}$\fi}
\newcommand{\pme}[2]{$^{+#1}_{-#2}$}
\newcommand{\pe}[1]{$^{+#1}$}
\newcommand{\me}[1]{$_{-#1}$}
\shorttitle{\textit{HST}/COS Observations of Outflows in 2MASS J1051+1247}
\shortauthors{Miller et al.}

\begin{document}

\title{\textit{HST}/COS Observations of Quasar Outflows in the 500 -- 1050 \AA\ Rest Frame: III\\Four Similar Outflows in 2MASS J1051+1247 with Enough Energy to be Major Contributors to AGN Feedback}
\altaffiliation{Based on observations with the NASA/ESA \\\textit{Hubble Space Telescope} obtained at the Space Telescope \\Science Institute, which is operated by the Association \\of Universities for Research in Astronomy, Incorporated, \\under NASA contract NAS5-26555.}

\author[0000-0002-0730-2322]{Timothy R. Miller}
\affiliation{Department of Physics, Virginia Polytechnic Institute and State University, Blacksburg, VA 24061, USA}
\author[0000-0003-2991-4618]{Nahum Arav}
\affiliation{Department of Physics, Virginia Polytechnic Institute and State University, Blacksburg, VA 24061, USA}
\author[0000-0002-9217-7051]{Xinfeng Xu}
\affiliation{Department of Physics, Virginia Polytechnic Institute and State University, Blacksburg, VA 24061, USA}
\author[0000-0002-2180-8266]{Gerard A. Kriss}
\affiliation{Space Telescope Science Institute, 3700 San Martin Drive, Baltimore, MD 21218, USA}
\author{Rachel J. Plesha}
\affiliation{Space Telescope Science Institute, 3700 San Martin Drive, Baltimore, MD 21218, USA}



\begin{abstract}
We detect four very energetic outflows in the \textit{Hubble Space Telescope}/Cosmic Origin Spectrograph spectra of quasar 2MASS J1051+1247 with a combined kinetic luminosity ($\dot{\sub{E}{K}}$) of 10$^{46}$~erg~s$^{-1}$. Remarkable similarities are seen in these outflows: velocity centroids between 4900 and 5700~km~s$^{-1}$, distances from the central source ($R$) of a few hundred parsecs that are all consistent within the errors, and an $\dot{\sub{E}{K}}$ within a factor of two for all outflows. Hence, a common origin for the outflows is probable. Most of the outflowing mass resides in a very high-ionization phase evident by troughs from \sion{Ne}{viii}, \sion{Na}{ix}, \sion{Mg}{x}, and \sion{Si}{xii}, which connect the physical conditions of these ultraviolet outflows to the X-ray warm absorber outflows seen in nearby Seyfert galaxies. Three of the outflows have two or three independent diagnostics for the electron number density, yielding consistent values for each outflow, which increase the robustness of the $R$ determinations. Troughs from never-before-seen ionic transitions of \sion{Ar}{vi}, \sion{O}{iv*}, \sion{Ne}{vi*}, and \sion{Ne}{v*} are identified. With a combined $\dot{\sub{E}{K}}$ that is 7.0\pme{6.5}{2.3}\% of the quasar's Eddington luminosity, these outflows are prime candidates to be major agents for various active galactic nuclei feedback effects.
\end{abstract}

\section{Introduction}\label{sec:int}
Blueshifted absorption troughs in the rest frame of quasar spectra are used to identify outflowing material from the host galaxy. A large fraction \cite[up to 40\%;][]{hew03,dai08,gan08,kni08} of the quasar population shows absorption outflows. Many feedback processes seen in active galactic nuclei (AGN) are likely caused by these outflows (see elaboration in section 1 of \cite{ara20a}, hereafter Paper I, and references therein).

The potential for these outflow systems to produce the aforementioned feedback rests primarily on their kinetic luminosity ($\dot{\sub{E}{K}}$), of which is linearly dependent on the distance from the central source ($R$). Simultaneously determining the electron number density (\sub{n}{e}) and ionization parameter (\sub{U}{H}) of the outflow is the most robust way to infer these distances \cite[see section 7.1 of][]{ara18}. Our group and others have used this method to publish around 20 such distances (see section 1 of Paper I and references therein). The range for these distances is between parsecs to tens of kiloparsecs and is orders of magnitude larger than theoretical predictions \cite[accretion disk wind models predict $\sim$0.03 pc; e.g.,][]{mur95,pro00,pro04}.

The ratio of the kinetic luminosity with respect to the Eddington luminosity is used to judge the feedback potential. Ratios exceeding 0.5\% \cite[]{hop10} or 5\% \cite[]{sca04} are thought to be sufficient. There are six known outflow systems that meet at least one of these criteria \cite[]{moe09,ara13,bor13,cha15b,xu19}.

The observations analyzed here were taken during Cycle 24 (GO-14777, PI: N. Arav) as part of a spectroscopic survey of 10 quasars with known outflows and redshifts around 1. The goal was to probe the 500-1050~\AA\ rest frame wavelength range (EUV500) for numerous diagnostic troughs like those listed in \cite{ara13} that can yield \sub{n}{e} and also troughs that arise from very high-ionization potential ions (e.g. \sion{Ne}{viii}, \sion{Mg}{x}, and \sion{Si}{xii}) that are typically seen in X-ray warm absorbers \cite[e.g.,][]{rey97,kaa00,cre03,kaa14}. With these very high-ionization potential ions, a connection can be established between X-ray warm absorbers and ultraviolet (UV) AGN outflows \cite[][]{ara13}.

This paper is part of a series of publications describing the results
of \textit{Hubble Space Telescope} (\textit{HST}) program GO-14777. 
\\
Paper I summarizes the results
for the individual objects and discusses their importance to various
aspects of quasar outflow research. 
\\
Paper II \cite[][]{xu20a} gives the full
analysis for four outflows detected in SDSS J1042+1646, including the
largest kinetic luminosity ($10^{47}$~erg~s$^{-1}$) outflow measured to date
at $R=800$~pc and another outflow at $R=15$~pc. 
\\
Paper III is this work.
\\
Paper IV \cite[][]{xu20b} presents the largest
velocity shift and acceleration measured to date in a broad absorption line (BAL) outflow.  
\\
Paper V \cite[][]{mil20b} analyzes two outflows
detected in PKS J0352-0711, including one outflow at $R=500$~pc and a second
outflow at $R=10$~pc that shows an ionization potential-dependent
velocity shift for troughs from different ions. 
\\
Paper VI \cite[][]{xu20c} analyzes two outflows
detected in SDSS J0755+2306, including one at $R=1600$~pc with
$\dot{E}_k=10^{46}-10^{47}$~erg~s$^{-1}$. 
\\
Paper VII (Miller et al. 2020c, in preparation) discusses the other
objects observed by program GO-14777, whose outflow characteristics
make the analysis more challenging.

The structure of this paper is as follows. Section~\ref{sec:od} describes the new observations of 2MASS J10512569+1247462 (hereafter, 2MASS J1051+1247) taken by the \textit{HST}/Cosmic Origins Spectrograph \cite[COS;][]{gre12}. The spectral fitting for the unabsorbed continuum and emission lines is also discussed. Determinations of the ionic column densities and electron number densities as well as the photoionization modeling are in Section~\ref{sec:da}. Section~\ref{sec:rd} presents our results on the physical properties, distances, and energetics of each outflow followed by a discussion in Section~\ref{sec:ds}. Section~\ref{sec:sc} closes with a summary and conclusions. Throughout this paper, we adopt a cosmology of $h = 0.696$, $\Omega_m = 0.286$, and $\Omega_\Lambda = 0.714$ and use Ned Wright's Javascript Cosmology Calculator website \cite[][]{wri06}.

\section{Observations, Data Reduction, and Spectral Fitting}
\label{sec:od}
\begin{figure*}
\includegraphics[scale=1.0]{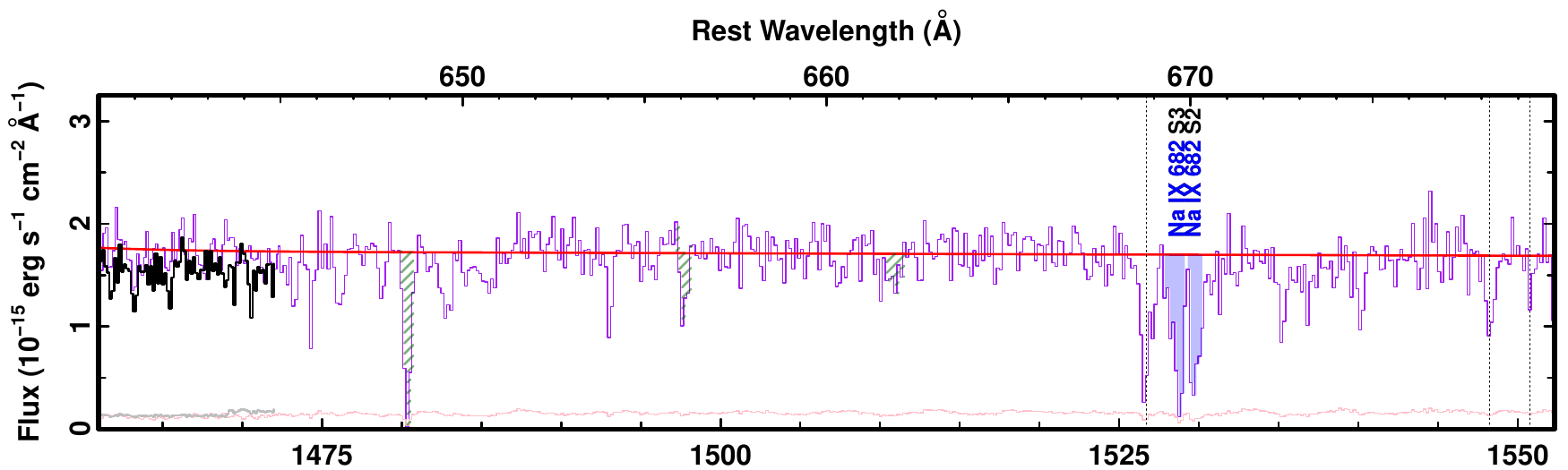}
\includegraphics[scale=1.0]{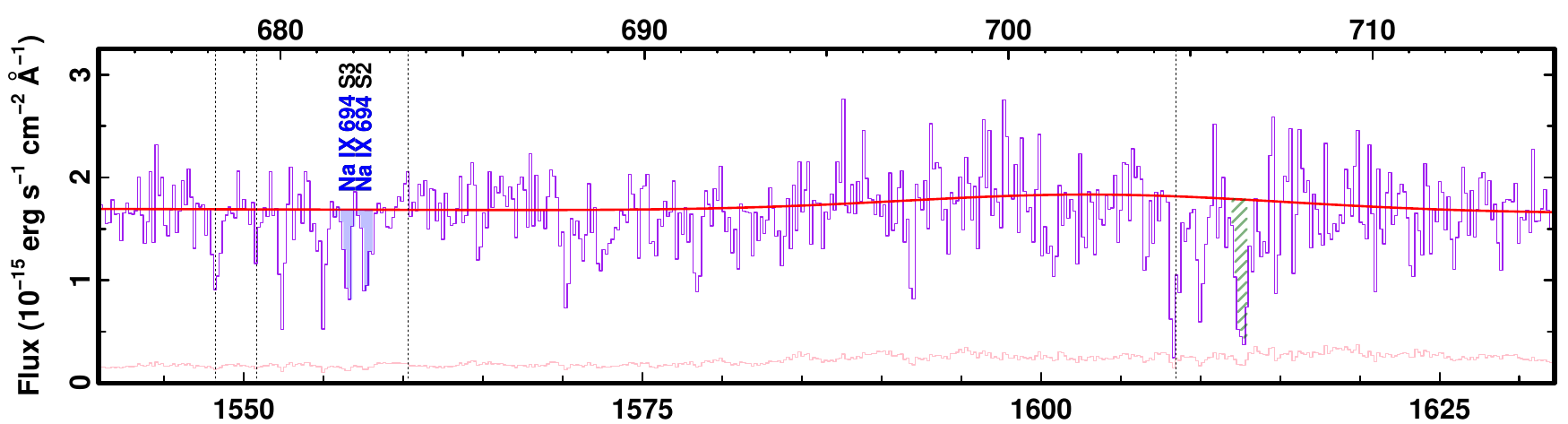}
\includegraphics[scale=1.0]{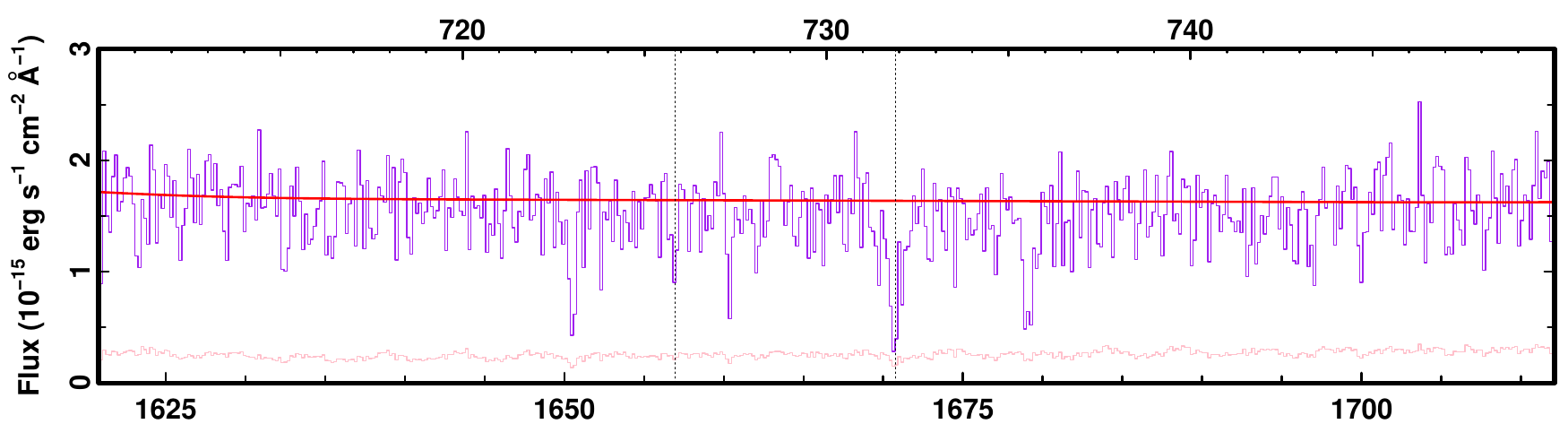}
\includegraphics[scale=1.0]{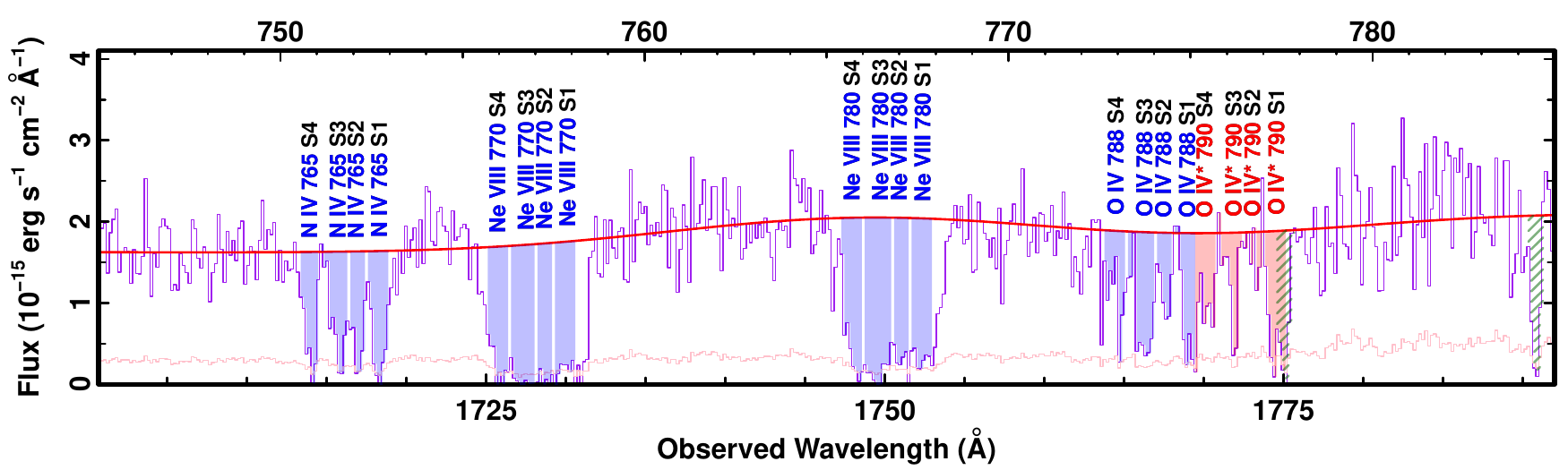}
\caption{\footnotesize{Dereddened, 2013 \textit{HST}/COS spectrum (in black) with errors (in gray) along with the dereddened 2018 spectrum (in purple) and errors (in light red). The main absorption troughs are labeled for all outflow systems (S1 = --4900~km~s$^{-1}$, S2 = --5150~km~s$^{-1}$, S3 = --5350~km~s$^{-1}$, S4 = --5650~km~s$^{-1}$). Identifications for transitions yielding upper limits are excluded, except for \sion{Ar}{vi*} 597 of S1 (see section 3.1). Blue shaded regions mark transitions from resonance absorption lines, and red regions mark excited ones. Blended troughs (B1--B10) are also labeled (see Section~\ref{sec:od}). Absorption troughs from intervening systems are the slanted dark green shaded regions, and the vertical dashed lines mark Galactic absorption and geocoronal emission features. The red contour traces the unabsorbed emission model for the 2013 data up to 645 \AA\ (rest frame) and the 2018 data at larger wavelengths.}}
\label{fig:spectrum}
\end{figure*}

2MASS J1051+1247 (J2000: R.A.~=~10:51:25.69, decl.~=~+12:47:46.2, $z$~=~1.2828) was first observed by \textit{HST}/COS in 2013 May (PID 12603) and again in 2018 January (PID 14777). Table~\ref{tab:obs} contains the details of each observation. Both datasets were processed in the same way as described in \cite{mil18} and were corrected for Galactic extinction with \textit{E}(\textit{B-V}) = 0.0202 \cite[]{sch11}. Figure~\ref{fig:spectrum} shows the dereddened, one-dimensional spectra in black and purple with errors in gray and light red for the 2013 and 2018 epochs, respectively. Absorption troughs for the four outflow systems are delineated S1, S2, S3, and S4 with centroid velocities and widths summarized in Table~\ref{tab:out}. All four outflows contain at least one previously undetected absorption trough: \sion{Ar}{vi} 544.73~\AA\ and 548.90~\AA, \sion{O}{IV*} 555.26~\AA, \sion{Ne}{vi*} 562.80~\AA, and \sion{Ne}{v*} 569.83~\AA\ and 572.34~\AA. Intervening hydrogen absorption systems are marked with slanted, dark green lines. The labels B1--B10 are the following blended troughs: B1 = \sion{O}{iv} 553 S2, \sion{O}{iv} 553 S3, and \sion{O}{iv} 554 S4; B2 = \sion{O}{iv} 553 S1, \sion{O}{iv} 554 S3, and \sion{O}{iv*} 554.5 S4; B3 = \sion{O}{iv} 554 S2 and \sion{O}{iv*} 554.5 S3; B4 = \sion{O}{iv} 554 S1, \sion{O}{iv*} 554.5 S2, and \sion{O}{iv*} 555 S4; B5 = \sion{O}{iv*} 554.5 S1 and \sion{O}{iv*} 555 S3; B6 = \sion{Ne}{v} 568 S1 and \sion{Ne}{v*} 570 S4; B7 = \sion{O}{iv} 608 S1, \sion{Mg}{x} 610 S4, and \sion{O}{iv*} 610 S4; B8 = \sion{Mg}{x} 610 S3 and \sion{O}{iv*} 610 S3; B9 = \sion{Mg}{x} 610 S2 and \sion{O}{iv*} 610 S2; B10 = \sion{Mg}{x} 610 S1 and \sion{O}{iv*} 610 S1.

Following the methodology of \cite{mil18}, the continuum emission was fitted with a power law, and line emission features were modeled with Gaussian profiles. The Gaussian fits were constrained by the red side of each line, avoiding the absorption that occurs mostly on the blue side of any given emission line. The Gaussian centroids were fixed at the rest frame wavelength of each emission line. The solid red contour in Figure~\ref{fig:spectrum} shows the unabsorbed emission model adopted in this work for the 2013 epoch up to 645 \AA\ (rest frame) and the 2018 epoch at larger wavelengths.
\begin{deluxetable}{lccc}
	\tablecaption{\textit{HST}/COS observations from 2013 to 2018 for 2MASS J1051+1247.\label{tab:obs}}
	\tablewidth{0pt}
	\tabletypesize{\footnotesize}
	\tablehead{
		& \multicolumn{3}{c}{Date}
	}
	\startdata
	& 2013 May 17 & 2018 Jan 4 & 2018 Jan 4 \\
	\tableline
	\textit{HST}/COS grating & G130M & G130M & G160M \\
	Exposure time (s) & 10,869 & 3460 & 4640 \\
	Observed range (\AA) & 1145--1470 & 1130--1470 & 1405--1800 \\
	Rest-frame range (\AA) & 500--645 & 495--645 & 615--790 \\
	\enddata
\end{deluxetable} 
\begin{deluxetable}{ccc}
	\tablecaption{Detected Outflows in 2MASS J1051+1247\label{tab:out}}
	\tablewidth{0pt}
	\tabletypesize{\footnotesize}
	\tablehead{
		\colhead{Outflow System} & \colhead{Centroid Velocity } & \colhead{FWHM}\\ 
		& \colhead{(km~s$^{-1}$)} & \colhead{(km~s$^{-1}$)} 
	}
	\startdata
	S1 & --4900 & 250\\
	S2 & --5150 & 200\\
	S3 & --5350 & 300\\
	S4 & --5650 & 250\\
	\enddata
	\end{deluxetable}

\begin{figure*}
\includegraphics[scale=1.0]{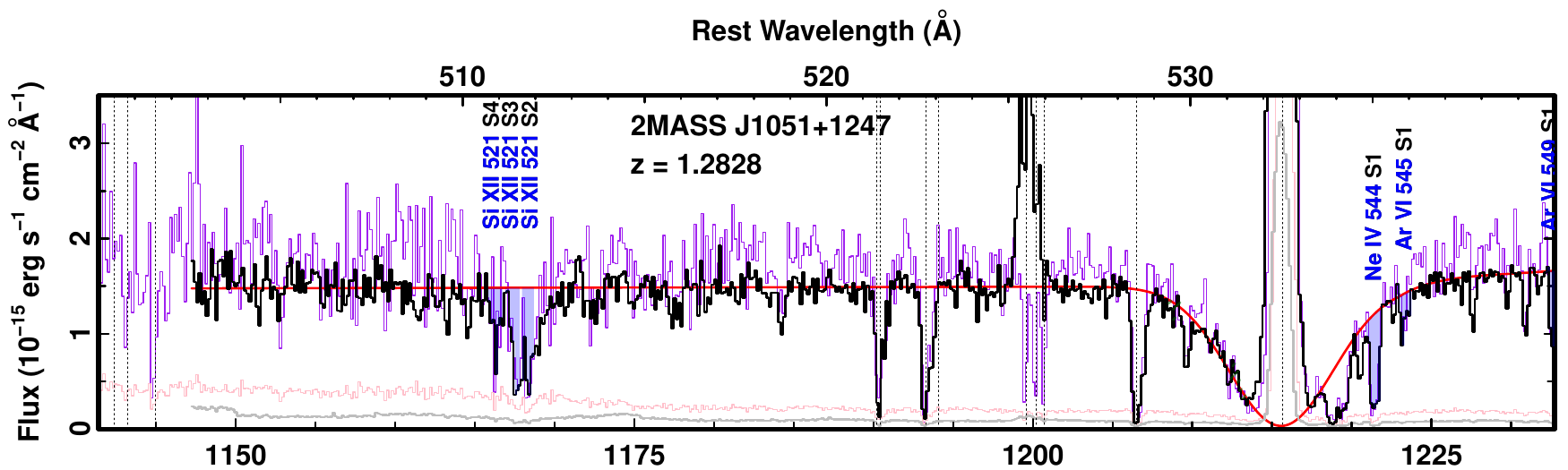}
\includegraphics[scale=1.0]{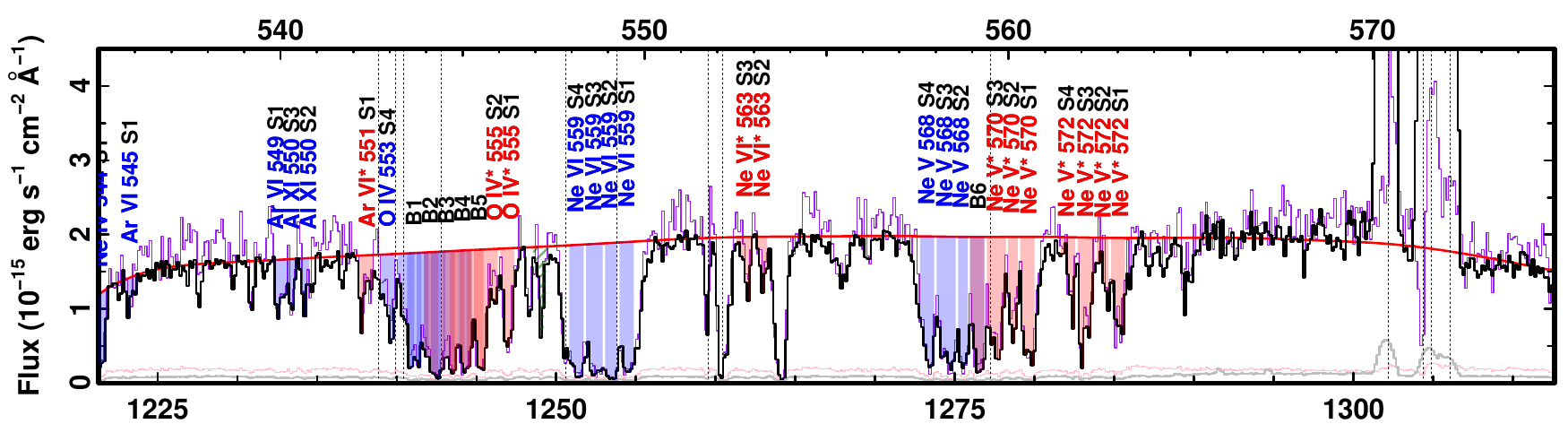}
\includegraphics[scale=1.0]{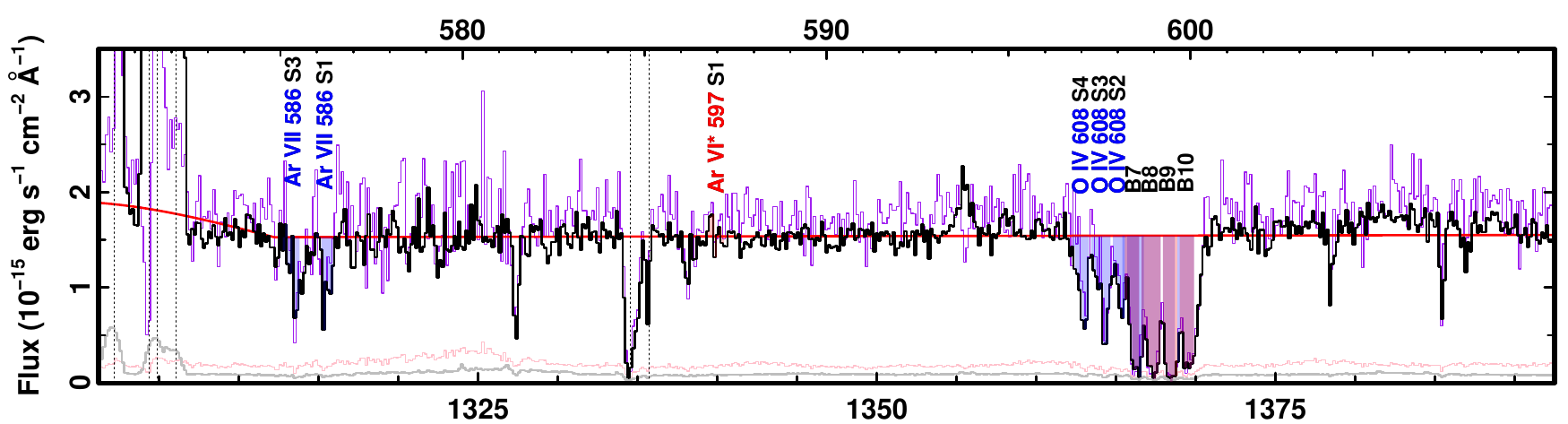}
\includegraphics[scale=1.0]{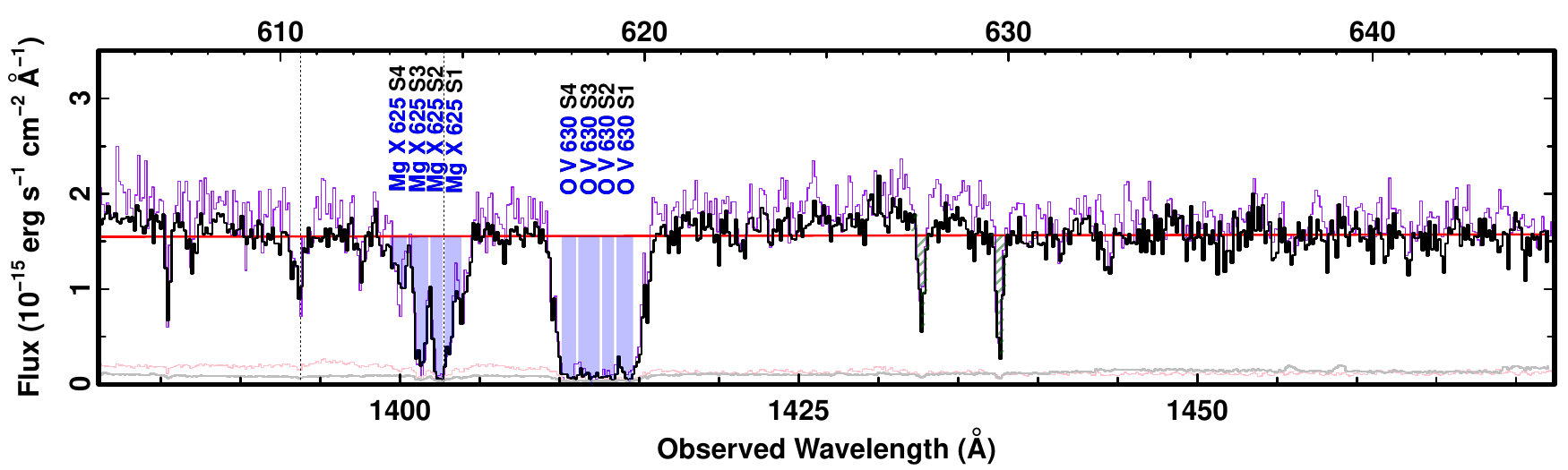}
\end{figure*}

\section{Data Analysis}
\label{sec:da}
\subsection{Ionic Column Density}
\label{sec:cd}
As detailed in \cite{mil18}, the apparent optical depth (AOD) and partial covering (PC) methods were used to measure ionic column densities (\sub{N}{ion}). The AOD method uses one ionic transition, while the PC method uses two ionic transitions to determine a single \sub{N}{ion} for each ionic energy state. The PC method is applicable when multiple lines (with different oscillator strengths, $f$) from the same ionic energy state have different trough depths, yielding a viable partial covering solution. Upon visual inspection, the troughs within the normalized spectra from each epoch did not show significant variability. Therefore, we used the column density measurements from the 2013 epoch when possible since the signal to noise is larger. 

The sum of all ionic energy state \sub{N}{ion} yields the total column density of each ion as listed in Table~\ref{tab:col} for all outflows. Upper and lower limits are highlighted in red and blue, respectively. S1 has an upper and lower limit \sub{N}{ion} for \sion{Ar}{vi}. The total \sub{N}{ion} for \sion{Ar}{vi} is the sum of the column densities for the resonance and excited states. The lower limit is obtained from the \sion{Ar}{vi} 548.90 and 544.73~\AA\ resonance troughs using the PC method without the addition of the excited state \sub{N}{ion}. The \sion{Ar}{vi*} 551.36~\AA\ trough is contaminated with unidentified absorption, so only an upper limit for \sion{Ar}{vi*} can be obtained from the \sion{Ar}{vi*} 596.67~\AA\ region (85\pe{30}$\times$10$^{12}$ cm$^{-2}$), yielding an overall upper limit to the total \sub{N}{ion} for \sion{Ar}{vi} as listed in the table. The last column contains the ratio of the adopted column densities to the best-fit, model predicted column densities (see Section~\ref{sec:photmod} and Figure~\ref{fig:sol}). This ratio is expected to be less than one for measured \sub{N}{ion} lower limits and vice versa for upper limits. The trough labels in Figure~\ref{fig:spectrum} combine multiple transitions with wavelength separations less than 0.5~\AA\ into a single transition. Table 3 of Paper II provides a list of transition atomic data. 

Using the same criteria of Paper II to account for non-black saturation, all PC determined \sub{N}{ion} are treated as measurements, \sub{N}{ion} measured for regions where no trough is identified (maximum optical depth, $\tau_{max}$, less than 0.05) are upper limits, and \sub{N}{ion} from troughs where both 0.05 $<$ $\tau_{max}$ $<$ 0.5 and troughs from ions of similar ionization potential that have $\tau_{max}$ $>$ 2 are also treated as measurements. Our adopted values are the PC values when available and AOD values otherwise. A systematic error, 20\% of the adopted value, is added in quadrature with the corresponding AOD/PC errors, yielding the adopted error values (see Table~\ref{tab:col}). This systematic error accounts for uncertainties in the unabsorbed emission model \cite[e.g.,][]{mil18,xu18}. For example, \sion{N}{IV} of S1 has an AOD \sub{N}{ion} = 310\pme{70}{40}. Since we are treating the adopted value as a lower limit, we calculate the lower error as $\sqrt{(40)^2+(0.2*310)^2} \approx 70$. 
\startlongtable
\begin{deluxetable}{ccccc}
\tablecaption{Total Ionic Column Densities\label{tab:col}}
\tablewidth{0pt}
\tabletypesize{\footnotesize}
\tablehead{
\colhead{Ion} & \colhead{AOD\tablenotemark{a}} & \colhead{PC\tablenotemark{a}} & \colhead{Adopted\tablenotemark{b}} & \colhead{\scriptsize$\frac{\textnormal{Adopted}}{\textnormal{Best Model}}$\tablenotemark{d}} \\ 
 & \colhead{($10^{12} $cm$^{-2}$)} & \colhead{($10^{12} $cm$^{-2}$)} & \colhead{($10^{12} $cm$^{-2}$)} & 
}
\startdata
\multicolumn{5}{c}{v = --4900~km~s$^{-1}$}\\
\tableline
\sion{N}{iv} & 310\pme{70}{40} & \nodata & \color{blue}\textgreater310\me{70} & \textgreater0.23\me{0.05}\\
\sion{O}{iii} & 520\pme{90}{70} & \nodata & \color{red}\textless520\pe{150} & \textless3.07\pe{0.89}\\
\sion{O}{iv} & 3100\pme{830}{280} & \nodata & \color{blue}\textgreater3100\me{670} & \textgreater0.30\me{0.07}\\
\sion{O}{v} & 820\pme{60}{40} & \nodata & \color{blue}\textgreater820\me{170} & \textgreater0.02\me{0.004}\\
\sion{Ne}{iv} & 980\pme{50}{50} & \nodata & \color{blue}\textgreater980\me{200} & \textgreater1.00\me{0.21}\\
\sion{Ne}{v} & 5900\pme{140}{70} & \nodata & \color{blue}\textgreater5900\me{1160} & \textgreater0.72\me{0.14}\\
\sion{Ne}{vi} & 3900\pme{130}{110} & \nodata & \color{blue}\textgreater3900\me{790} & \textgreater0.23\me{0.05}\\
\sion{Ne}{viii} & 4600\pme{770}{470} & \nodata & \color{blue}\textgreater4600\me{1100} & \textgreater0.20\me{0.04}\\
\sion{Na}{ix} & 240\pme{110}{110} & \nodata & \color{red}\textless240\pe{120} & \textless0.98\pe{0.49}\\
\sion{Mg}{x} & 3300\pme{150}{140} & \nodata & \color{blue}\textgreater3300\me{670} & \textgreater0.94\me{0.19}\\
\sion{Al}{xi} & 170\pme{50}{50} & \nodata & \color{red}\textless170\pe{50} & \textless3.36\pe{0.99}\\
\sion{Si}{xii} & 1600\pme{230}{180} & \nodata & \color{red}\textless1600\pe{410} & \textless26.77\pe{6.44}\\
\sion{S}{iv} & 25\pme{4}{4} & \nodata & \color{red}\textless25\pe{7} & \textless1.00\pe{0.28}\\
\sion{Cl}{vi} & 9.5\pme{4.2}{3.5} & \nodata & \color{red}\textless9.5\pe{4.6} & \textless1.07\pe{0.52}\\
\sion{Ar}{vi} & 130\pme{20}{20} & 160\pme{20}{20} & \color{blue}\textgreater160\me{40} & \textgreater0.88\me{0.22}\\
\sion{Ar}{vi} & \nodata & \nodata & \color{red}\textless250\pe{70} & \textless1.37\pe{0.38}\\
\sion{Ar}{vii} & 44\pme{4}{3} & \nodata & \color{blue}\textgreater44\me{9} & \textgreater1.04\me{0.21}\\
\sion{Ar}{viii} & 120\pme{40}{30} & \nodata & \color{red}\textless120\pe{50} & \textless1.73\pe{0.72}\\
\sion{Ca}{vi} & 180\pme{80}{60} & \nodata & \color{red}\textless180\pe{90} & \textless0.91\pe{0.45}\\
\sion{Ca}{viii} & 130\pme{50}{40} & \nodata & \color{red}\textless130\pe{50} & \textless0.95\pe{0.36}\\
\tableline
\multicolumn{5}{c}{v = --5150~km~s$^{-1}$}\\
\tableline
\sion{N}{iv} & 210\pme{80}{30} & \nodata & \color{blue}\textgreater210\me{50} & \textgreater0.53\me{0.13}\\
\sion{O}{iii} & 250\pme{60}{50} & \nodata & \color{red}\textless250\pe{100} & \textless6.69\pe{2.68}\\
\sion{O}{iv} & 1400\pme{100}{170} & \nodata & \color{blue}\textgreater1400\me{350} & \textgreater0.45\me{0.10}\\
\sion{O}{v} & 810\pme{210}{40} & \nodata & \color{blue}\textgreater810\me{170} & \textgreater0.06\me{0.01}\\
\sion{Ne}{iv} & 270\pme{30}{30} & \nodata & \color{red}\textless270\pe{60} & \textless0.95\pe{0.21}\\
\sion{Ne}{v} & 3500\pme{110}{40} & \nodata & \color{blue}\textgreater3500\me{680} & \textgreater1.16\me{0.23}\\
\sion{Ne}{vi} & 5100\pme{230}{200} & \nodata & \color{blue}\textgreater5100\me{1100} & \textgreater0.82\me{0.17}\\
\sion{Ne}{viii} & 4200\pme{1000}{420} & \nodata & \color{blue}\textgreater4200\me{930} & \textgreater0.16\me{0.04}\\
\sion{Na}{ix} & 800\pme{140}{120} & 970\pme{80}{80} & 970\pme{210}{210} & 1.15\pme{0.25}{0.25}\\
\sion{Mg}{x} & 7700\pme{740}{510} & \nodata & \color{blue}\textgreater7700\me{1600} & \textgreater0.40\me{0.08}\\
\sion{Al}{xi} & 590\pme{40}{40} & \nodata & 590\pme{130}{130} & 0.76\pme{0.17}{0.17}\\
\sion{Si}{xii} & 3800\pme{310}{210} & \nodata & \color{blue}\textgreater3800\me{770} & \textgreater1.21\me{0.25}\\
\sion{S}{iv} & 7.2\pme{2.0}{3.4} & \nodata & \color{red}\textless7.2\pe{2.4} & \textless1.35\pe{0.45}\\
\sion{Cl}{vi} & 7.9\pme{2.8}{3.6} & \nodata & \color{red}\textless7.9\pe{3.2} & \textless2.59\pe{1.05}\\
\sion{Ar}{vi} & 90\pme{10}{10} & \nodata & \color{red}\textless90\pe{20} & \textless1.19\pe{0.26}\\
\sion{Ar}{vii} & 17\pme{3}{3} & \nodata & \color{red}\textless17\pe{5} & \textless0.89\pe{0.26}\\
\sion{Ar}{viii} & 65\pme{30}{20} & \nodata & \color{red}\textless65\pe{40} & \textless2.26\pe{1.39}\\
\sion{Ca}{vi} & 180\pme{60}{70} & \nodata & \color{red}\textless180\pe{70} & \textless2.28\pe{0.89}\\
\sion{Ca}{viii} & 200\pme{40}{40} & \nodata & \color{red}\textless200\pe{60} & \textless8.48\pe{2.54}\\
\tableline
\multicolumn{5}{c}{v = --5350~km~s$^{-1}$}\\
\tableline
\sion{N}{iv} & 250\pme{60}{30} & \nodata & \color{blue}\textgreater250\me{60} & \textgreater0.67\me{0.16}\\
\sion{O}{iii} & 560\pme{80}{80} & \nodata & \color{red}\textless560\pe{160} & \textless48.93\pe{13.98}\\
\sion{O}{iv} & 2100\pme{170}{200} & \nodata & \color{blue}\textgreater2100\me{480} & \textgreater0.93\me{0.20}\\
\sion{O}{v} & 1000\pme{110}{40} & \nodata & \color{blue}\textgreater1000\me{170} & \textgreater0.04\me{0.01}\\
\sion{Ne}{v} & 5800\pme{100}{90} & \nodata & \color{blue}\textgreater5800\me{1200} & \textgreater1.05\me{0.21}\\
\sion{Ne}{vi} & 5600\pme{130}{180} & \nodata & \color{blue}\textgreater5600\me{1200} & \textgreater0.29\me{0.06}\\
\sion{Ne}{viii} & 7800\pme{2200}{900} & \nodata & \color{blue}\textgreater7800\me{1800} & \textgreater0.26\me{0.06}\\
\sion{Na}{ix} & 760\pme{130}{100} & 1000\pme{80}{90} & 1000\pme{220}{220} & 1.08\pme{0.24}{0.24}\\
\sion{Mg}{x} & 4600\pme{230}{180} & \nodata & \color{blue}\textgreater4600\me{930} & \textgreater0.20\me{0.04}\\
\sion{Al}{xi} & 440\pme{40}{40} & \nodata & 440\pme{100}{100} & 0.62\pme{0.14}{0.14}\\
\sion{Si}{xii} & 4400\pme{270}{300} & \nodata & \color{blue}\textgreater4400\me{950} & \textgreater1.49\me{0.31}\\
\sion{S}{iv} & 35\pme{5}{5} & \nodata & \color{red}\textless35\pe{9} & \textless24.20\pe{6.22}\\
\sion{Cl}{vi} & 6.5\pme{2.8}{2.9} & \nodata & \color{red}\textless6.5\pe{3.1} & \textless2.30\pe{1.10}\\
\sion{Ar}{vi} & 110\pme{10}{10} & \nodata & \color{red}\textless110\pe{30} & \textless1.04\pe{0.28}\\
\sion{Ar}{vii} & 46\pme{3}{3} & \nodata & \color{blue}\textgreater46\me{10} & \textgreater0.76\me{0.16}\\
\sion{Ar}{viii} & 240\pme{50}{50} & \nodata & \color{red}\textless240\pe{70} & \textless1.87\pe{0.55}\\
\sion{Ca}{vi} & 150\pme{80}{60} & \nodata & \color{red}\textless150\pe{90} & \textless0.86\pe{0.52}\\
\sion{Ca}{viii} & 260\pme{40}{90} & \nodata & \color{red}\textless260\pe{70} & \textless1.73\pe{0.46}\\
\tableline
\multicolumn{5}{c}{v = --5650~km~s$^{-1}$}\\
\tableline
\sion{N}{iv} & 230\pme{70}{30} & \nodata & \color{blue}\textgreater230\me{60} & \textgreater0.45\me{0.12}\\
\sion{O}{iii} & 300\pme{70}{60} & \nodata & \color{red}\textless300\pe{110} & \textless3.82\pe{1.40}\\
\sion{O}{iv} & 1700\pme{220}{190} & \nodata & \color{blue}\textgreater1700\me{400} & \textgreater0.41\me{0.09}\\
\sion{O}{v} & 880\pme{60}{40} & \nodata & \color{blue}\textgreater880\me{180} & \textgreater0.07\me{0.01}\\
\sion{Ne}{v} & 2700\pme{100}{40} & \nodata & \color{blue}\textgreater2700\me{520} & \textgreater1.01\me{0.20}\\
\sion{Ne}{vi} & 4600\pme{160}{160} & \nodata & \color{blue}\textgreater4600\me{950} & \textgreater1.11\me{0.23}\\
\sion{Ne}{viii} & 5000\pme{1100}{600} & \nodata & \color{blue}\textgreater5000\me{1200} & \textgreater0.27\me{0.06}\\
\sion{Na}{ix} & 900\pme{140}{130} & \nodata & \color{red}\textless900\pe{220} & \textless1.69\pe{0.41}\\
\sion{Mg}{x} & 1000\pme{120}{80} & \nodata & \color{blue}\textgreater1000\me{210} & \textgreater0.09\me{0.02}\\
\sion{Al}{xi} & 330\pme{50}{40} & \nodata & \color{red}\textless330\pe{80} & \textless0.80\pe{0.20}\\
\sion{Si}{xii} & 1800\pme{250}{180} & \nodata & \color{blue}\textgreater1800\me{390} & \textgreater1.23\me{0.28}\\
\sion{S}{iv} & 12\pme{4}{4} & \nodata & \color{red}\textless12\pe{5} & \textless0.98\pe{0.41}\\
\sion{Cl}{vi} & 5.7\pme{3.3}{2.5} & \nodata & \color{red}\textless5.7\pe{3.5} & \textless1.80\pe{1.10}\\
\sion{Ar}{vi} & 70\pme{10}{9} & \nodata & \color{red}\textless70\pe{20} & \textless1.21\pe{0.34}\\
\sion{Ar}{vii} & 10\pme{3}{2} & \nodata & \color{red}\textless10\pe{3} & \textless0.95\pe{0.29}\\
\sion{Ar}{viii} & 60\pme{30}{30} & \nodata & \color{red}\textless60\pe{30} & \textless4.22\pe{2.11}\\
\sion{Ca}{vi} & 190\pme{90}{90} & \nodata & \color{red}\textless190\pe{90} & \textless2.97\pe{1.40}\\
\sion{Ca}{viii} & 430\pme{90}{180} & \nodata & \color{red}\textless430\pe{90} & \textless31.06\pe{6.50}\\
\enddata
\tablenotetext{a}{Sum of all \sub{N}{ion} from excited and resonance states for a given ion in \\each outflow system using the AOD and PC methods.}
\tablenotetext{b}{The adopted values in blue are lower limits, in red are upper limits, \\and in black are measurements.}
\tablenotetext{c}{The ratio of the adopted values to the column densities from the \\best-fit Cloudy model.}
\end{deluxetable}

\subsection{Photoionization Modeling}
\label{sec:photmod}
Since the troughs in each outflow are narrow and the blended troughs are not critical to the analysis, we do not use the Synthetic Spectral Simulation (SSS) method presented in Paper II and instead follow the methodology of prior works \cite[e.g.,][]{mil18,xu18,xu19}. Each outflow system is modeled with a hydrogen column density (\sub{N}{H}) and \sub{U}{H}. We generated grids of photoionization models with the code Cloudy \cite[][version c17.00]{fer17}. Each grid assumed one metallicity (two total) and one spectral energy distribution (SED, three total). The three SEDs are the UV-soft SED \cite[]{dun10}, the HE0238 SED \cite[]{ara13}, and the MF87 SED \cite[]{mat87} of which are a representative range of SED shapes that are applicable for radio-quiet quasars \cite[]{ara13}. The two metallicities are solar, Z$_{\astrosun}$, from \citet[][]{gre10} and super-solar, Z = 4.68Z$_{\astrosun}$, from Paper V. These parameters directly determine the model \sub{N}{ion}.

\begin{figure*}
\includegraphics[trim=5mm 24mm 4mm 16mm,clip,scale=0.354]{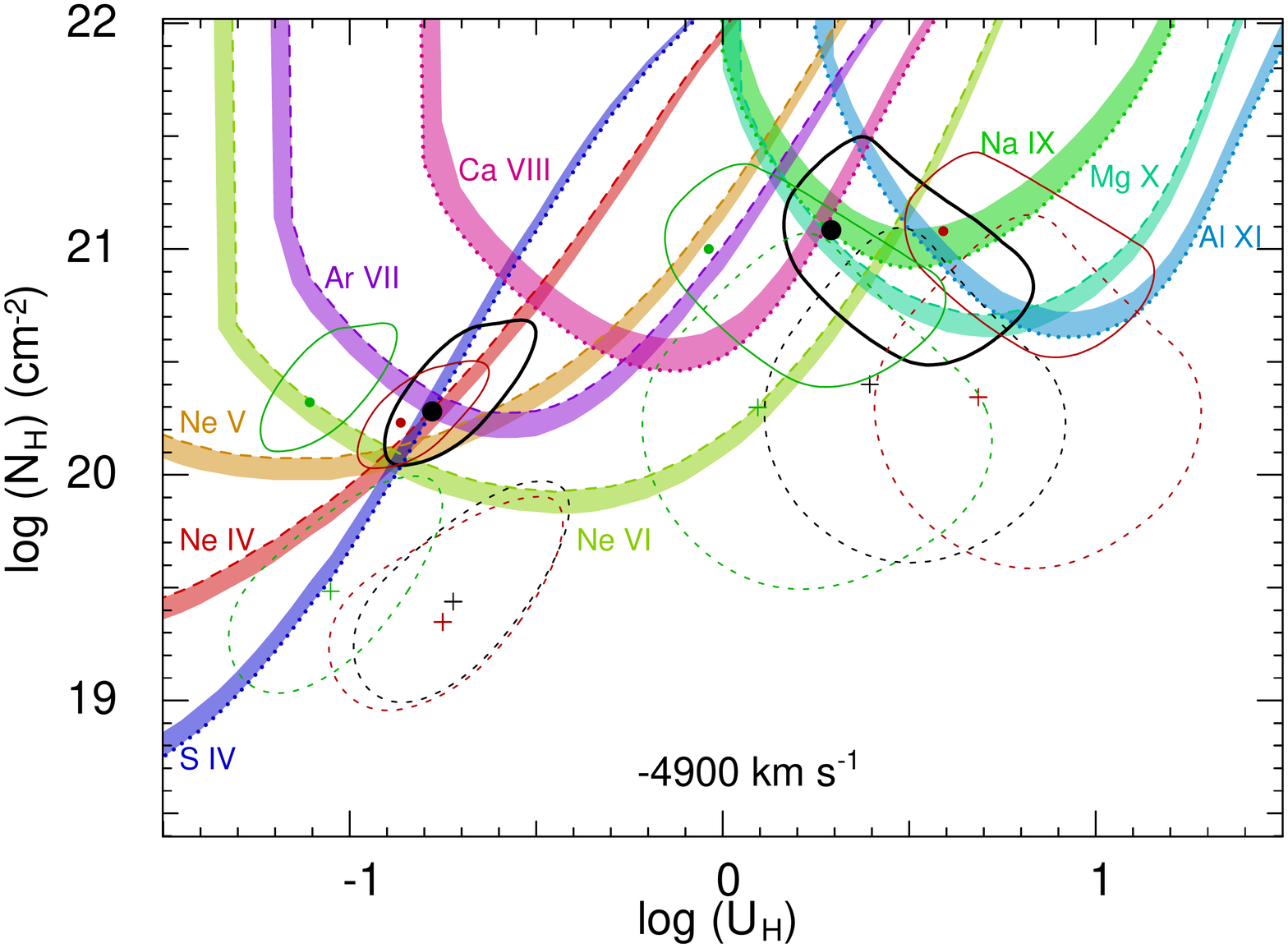}\includegraphics[trim=37mm 24mm 4mm 16mm,clip,scale=0.354]{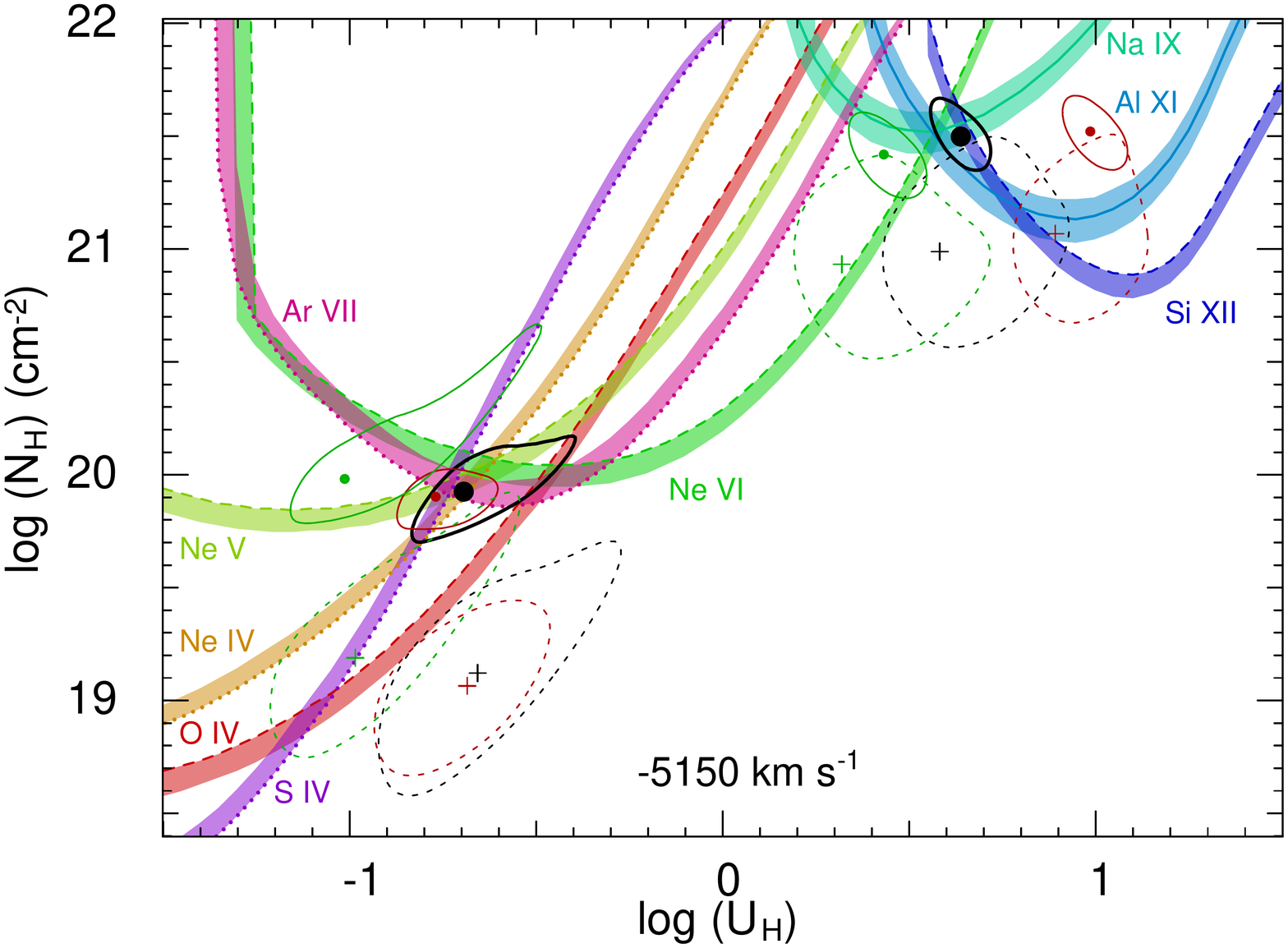}
\includegraphics[trim=5mm 1mm 4mm 17mm,clip,scale=0.354]{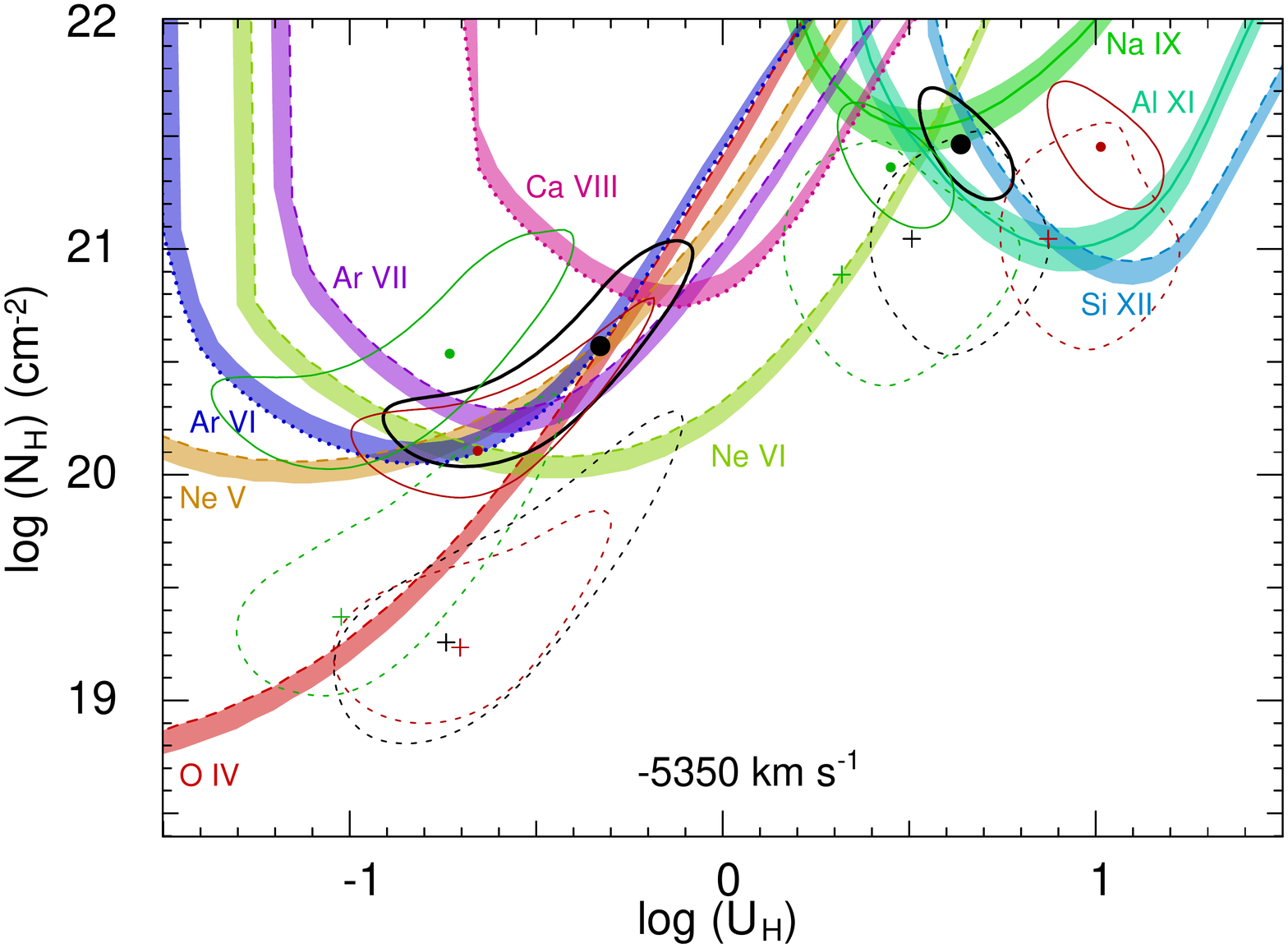}\includegraphics[trim=37mm 1mm 4mm 17mm,clip,scale=0.354]{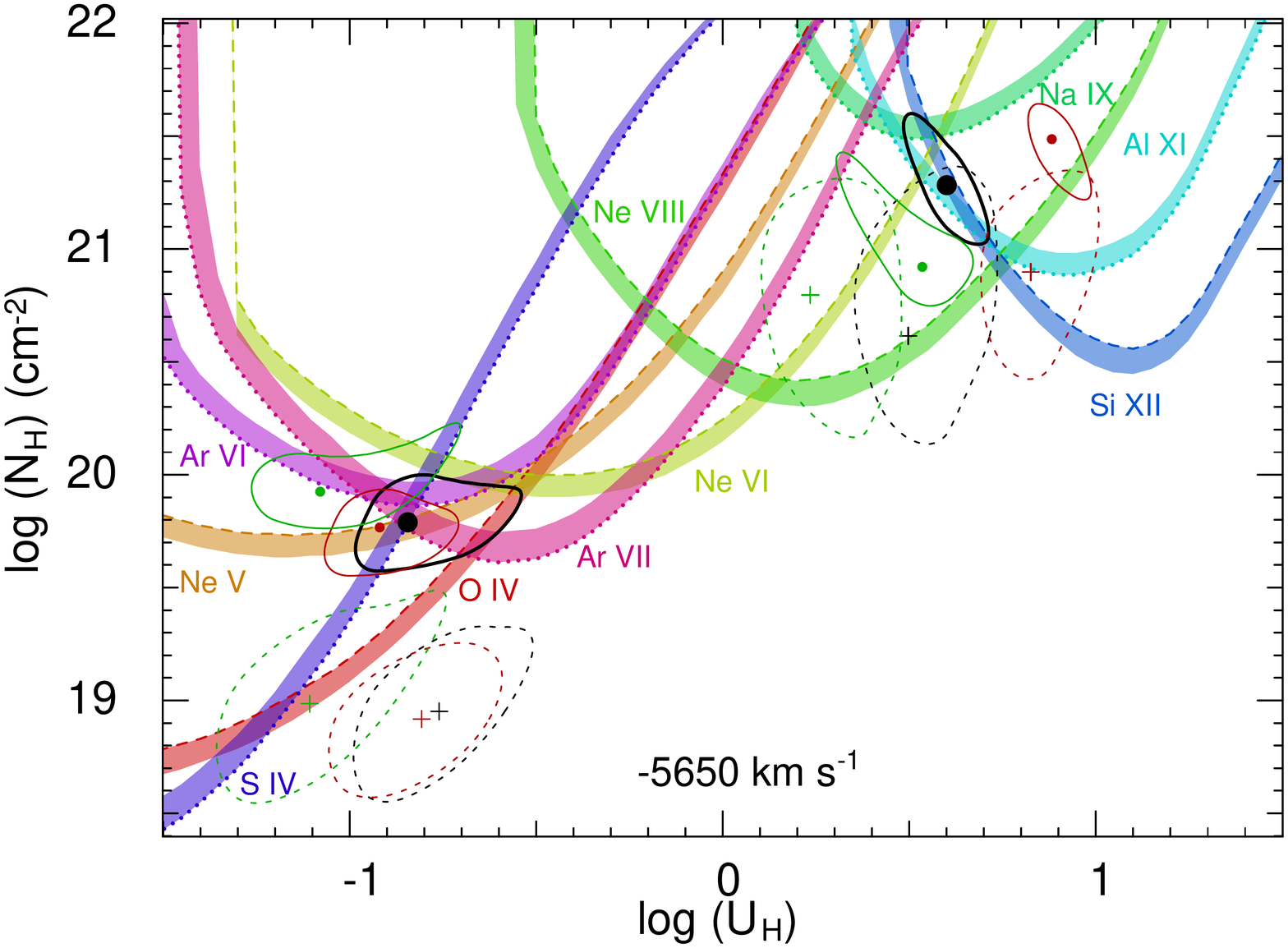}
\caption{\footnotesize{Two-phase photoionization solution for each outflow system. The colored contours show the model parameters that are consistent with the observed values assuming the HE0238 SED and solar metallicity. Solid contours represent ionic column densities taken as measurements, while dotted and dashed contours are upper and lower limits, respectively. The shaded bands are the 1$\sigma$ uncertainties for each contour (see Table~\ref{tab:col}). The dots are the best $\chi^2$-minimization solutions, assuming the solar metallicity for each ionization phase and the ellipses encircling them are their 1$\sigma$ uncertainties. The black, red, and green solutions are for the HE0238 SED, MF87 SED, and UV-soft SED, respectively. The plus symbol solutions for each phase assume Z = 4.68 Z$_{\astrosun}$} from Table~3 of Paper V and also account for the same uncertainty in metallicity.}
\label{fig:sol}
\end{figure*}

To determine the best pair of \sub{N}{H} and \sub{U}{H}, the measured \sub{N}{ion} are compared to the modeled values. The colored contours for individual ions in Figure~\ref{fig:sol} show the \sub{N}{H} and \sub{U}{H} pairs where the model \sub{N}{ion} are within 1$\sigma$ of the observed values, assuming the HE0238 SED and the solar metallicity. Solid contours represent \sub{N}{ion} measurements while dotted and dashed lines indicate upper and lower limits, respectively. $\chi^2$-minimization of the model \sub{N}{ion} compared to the measured \sub{N}{ion} from Table~\ref{tab:col} determines the best-fit solution. The adopted, best-fit solution is the HE0238 SED with the solar metallicity (solid black dots and 1$\sigma$ error ellipses). Assuming the solar metallicity and changing the SED results in the solid red (UV-soft SED) and solid green (MF87 SED) solutions. As expected, the \sub{N}{ion} contours shift according to the SED shape. For example, the UV-soft SED has a higher luminosity at the wavelengths needed to produce the high- and very high-ionization potential ions, resulting in a lower-ionization parameter in both phases. Similarly, assuming the super-solar metallicity decreases the hydrogen column density required to match the observations, and the associated solutions for each SED are the plus symbols with dashed ellipses.  

A two-phase photoionization solution \cite[][]{ara13} is needed for all outflow systems to satisfy the column density measurements from both the very high-ionization potential ions (e.g., \sion{Mg}{x}, \sion{Na}{ix}, and \sion{Al}{xi}) and high-ionization potential ions (e.g., \sion{Ne}{iv}, \sion{Ne}{v}, and \sion{O}{iv}). For S1, a single phase solution at the intersection of the \sion{Ne}{iv} and \sion{Na}{ix} contours over predicts the upper limit column densities of \sion{Ca}{vi}, \sion{Ca}{viii}, and \sion{Ar}{viii} by over an order of magnitude. Similar over predictions occur for the other outflow systems when a single phase solution is chosen. The values for all \sub{N}{H} and \sub{U}{H} determinations are given in Table~\ref{tab:res}.

\subsection{Electron Number Density Determination}
\label{sec:ed}
The excited state troughs shown in Figure~\ref{fig:spectrum} all become populated through electron collisions. The frequency of collisions and amount of energy transferred between bound and free electrons depend on \sub{n}{e} and the electron temperature. Therefore, \sub{n}{e} can be calculated from the relative populations between either two different excited states or an excited state and a resonance state from the same ion \cite[e.g.,][]{dek01,ham01,dek02,kor08}. We used the CHIANTI 8.0.7 database \cite[]{der97,lan13} to calculate the necessary population ratios (equal to the column density ratios) as was done in previous works \cite[e.g.,][]{bor12b,ara13,ara15,ara18,cha15b}.  

However, some of the observed excited states cannot be used with this method. For each outflow system, the \sion{O}{iv*} 790.20~\AA\ and 555.26~\AA\ troughs (different $f\cdot\lambda$) exhibit 1:1 trough depths (indicative of non-black saturation) with not only each other but also with the resonance transitions, making column density measurements lower limits for every trough. Therefore, the ratios are unconstrained. 

This leaves the \sion{Ne}{vi*} 562.81~\AA\ and \sion{Ne}{v*} 569.83~\AA\ and 572.34~\AA\ excited troughs with the \sion{Ne}{vi} 558.60~\AA\ and \sion{Ne}{v} 568.42~\AA\ resonance troughs as potentially useful density diagnostics. Since \sion{Ne}{vi} 558.60~\AA\ and \sion{Ne}{v} 568.42~\AA\ are the only resonance transitions for each ion and, therefore, yield lower limit measurements, we use the photoionization solutions to constrain the total column densities for these ions. Nearly all of the \sion{Ne}{v} column density in each outflow is produced by their respective high-ionization phases. Therefore, an upper limit to the \sub{N}{ion} for \sion{Ne}{v} is determined by finding the largest \sub{N}{ion} value contained within the 1$\sigma$ error ellipse of each high-ionization phase. When the photoionization solution yielded a model \sub{N}{ion} for \sion{Ne}{v} larger than the measured lower limit value, as is the case for S1, we chose the model \sub{N}{ion} for the ratio calculation. The lower limit \sub{N}{ion} for \sion{Ne}{v} was taken to be the measured lower limit value since it provided a tighter constraint compared to the smallest value obtained from the 1$\sigma$ error ellipse. 

The same process was done for calculating the \sub{N}{ion} and errors for \sion{Ne}{vi} in S3. The \sion{Ne}{vi} column densities for the other outflows are produced in roughly equal amounts from both phases. Assuming the phases are cospatial given the velocity correspondence, they have different \sub{n}{e} values. Therefore, the density cannot be reliably determined from those \sion{Ne}{vi} troughs since we would have to deconvolve the troughs into the separate phases. The excited state troughs are all shallower than their resonance counterparts. Therefore, assuming they have the same velocity-dependent covering factors, C(v), and using the maximum values, \sub{C}{max}(v) = 1-\sub{I}{res}(v) where \sub{I}{res}(v) is the velocity-dependent normalized flux for the resonance trough, the measurements of the excited state column densities are within 30\% of their true values, yielding usable density ratios.

\begin{figure}
\includegraphics[trim=6mm 9mm 9mm 5mm,clip,scale=0.35]{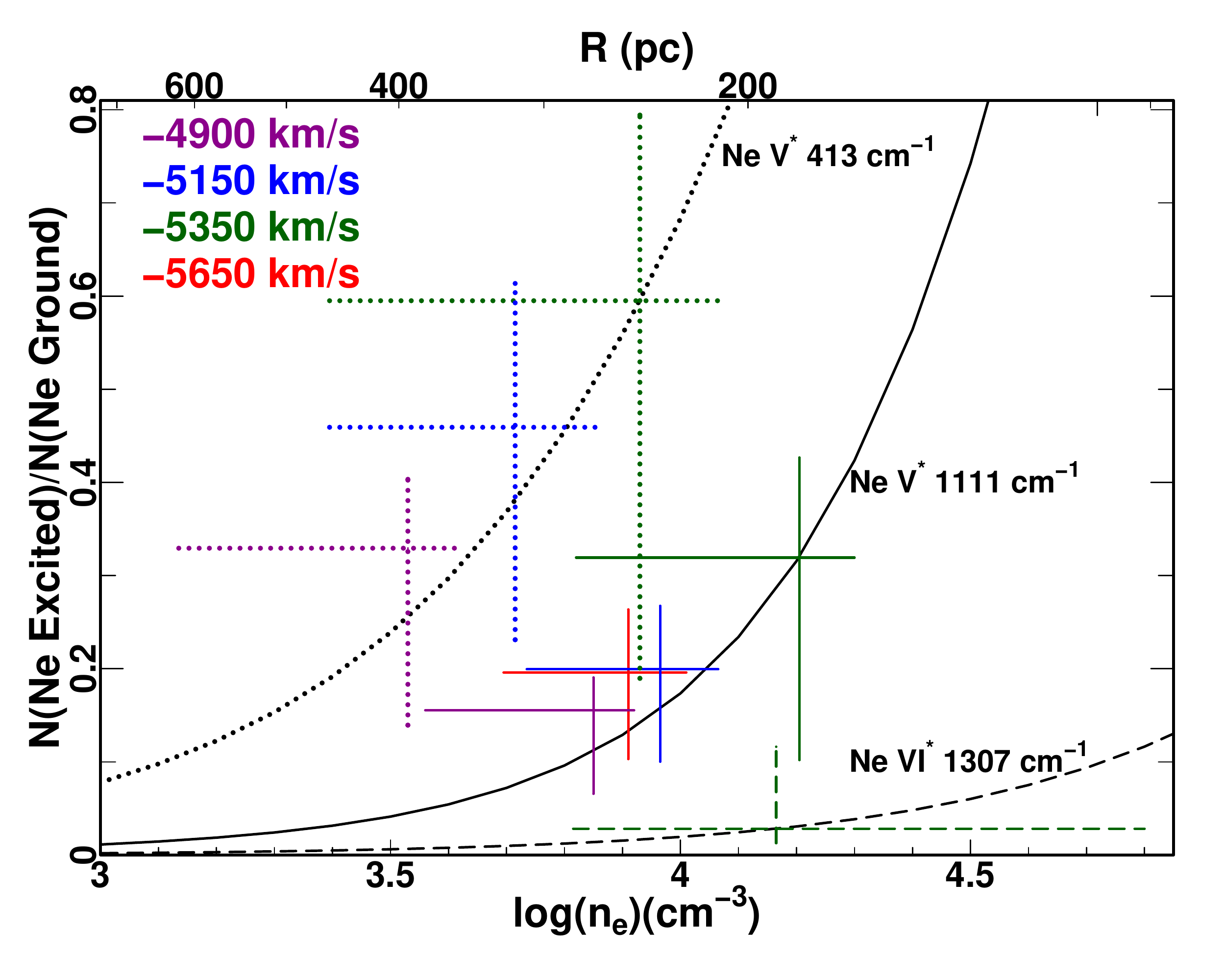}\\
\caption{\footnotesize{Electron number density, \sub{n}{e}, of each outflow system based on three population ratios of Ne. The theoretical predictions from CHIANTI for the population ratios with excited energy levels of \sion{Ne}{v*} 413~cm$^{-1}$, \sion{Ne}{v*} 1111~cm$^{-1}$, and \sion{Ne}{vi*} 1307~cm$^{-1}$ are overlaid. The curves assume the average temperature, 27,500 K, from the photoionization solution for the high-ionization phase of the --5350~km~s$^{-1}$ outflow. The corresponding distance, $R$ (from equation \ref{eq:R}), for this outflow is also shown on the top axis. The offset of the ratios from the shown curves for the other outflows are the result of different electron temperatures given by Cloudy for those outflow systems.}}
\label{fig:dens}
\end{figure}

In Figure~\ref{fig:dens}, the theoretical column density ratios as a function of \sub{n}{e} are shown with the black contours for each population ratio: dotted = N(\sion{Ne}{v*} 569.83~\AA)/N(\sion{Ne}{v} 568.42~\AA) = N(\sion{Ne}{v} 413~cm$^{-1}$)/N(\sion{Ne}{v} 0~cm$^{-1}$), solid = N(\sion{Ne}{v*} 572.34~\AA)/N(\sion{Ne}{v} 568.42~\AA) = N(\sion{Ne}{v} 1111~cm$^{-1}$)/N(\sion{Ne}{v} 0~cm$^{-1}$), and dashed = N(\sion{Ne}{vi*} 562.81~\AA)/N(\sion{Ne}{vi} 558.60~\AA) = N(\sion{Ne}{vi} 1307~cm$^{-1}$)/N(\sion{Ne}{vi} 0~cm$^{-1}$). The measured column density ratios with uncertainties for each outflow system are overlaid. The different ratios yield consistent \sub{n}{e} within the measurement errors for each outflow system. Therefore, we adopt the N(\sion{Ne}{v} 1111~cm$^{-1}$)/N(\sion{Ne}{v} 0~cm$^{-1}$) \sub{n}{e} values for the high-ionization phases since the \sion{Ne}{v*} 572.34~\AA\ troughs were the shallowest for each outflow system, yielding the most accurate \sub{N}{ion} determinations. The errors for \sub{n}{e} are determined by the horizontal intersection of the ratio including the errors with the CHIANTI curve, e.g., the \sub{n}{e} value where the CHIANTI curve gives the ratio adding the plus error yields the upper error bound on \sub{n}{e}. The \sub{n}{e} for the very high-ionization phases (VHP) are calculated by assuming the two phases are cospatial, and both \sub{n}{e} values are listed in Table~\ref{tab:res}.

\section{Results}
\label{sec:rd}
\subsection{Outflow Distance, Energetics, and Properties}
The distance each outflow is from the central source can be calculated from the definition of the ionization parameter: 
\begin{equation}
\label{eq:R}
\sub{U}{H} = \frac{\sub{Q}{H}}{4\pi R^2\ssub{n}{H} c}
\end{equation}
where \ssub{n}{H}~is the hydrogen number density with $\sub{n}{e} \approx 1.2\ssub{n}{H}$ for highly ionized plasma, $R$ is the distance from the central source, $c$ is the speed of light, and $\sub{Q}{H}$ is the incident ionizing photon rate of hydrogen. Integrating the HE0238 SED for energies above 1 Ryd yields $\sub{Q}{H}=7.3\times 10^{56}$~s$^{-1}$.

The distances of the outflows are given in Table~\ref{tab:res}, and they are consistent with being located at the same distance within the errors. The small separations in velocity between each outflow also suggests that they are connected. The full width of the \sion{Ne}{viii} trough across all outflows is $\sim$1200~km~s$^{-1}$. Using the classification scheme in section 4.5 of Paper I, the outflows as a whole are a mini-BAL. 

Assuming a partially filled thin shell outflow \cite[see section 5.3 here and][]{bor12}, the average mass flow rate and kinetic luminosity over the dynamical timescale ($R$/$v$) are given by
\begin{equation}
\label{eq:M}
\dot{M}\simeq 4\pi \Omega R N_H \mu m_p v 
\end{equation}
and
\begin{equation}
\label{eq:E}
\dot{\sub{E}{K}}\simeq \frac{1}{2} \dot{M} v^2
\end{equation}
where $\Omega$ = 0.4\pme{0.14}{0.14} is the global covering factor \cite[a fraction of quasars with observed \sion{Ne}{viii} mini-BAL outflows;][]{muz13}, $R$ is the distance from the central source, $\mu = 1.4$ is the mean atomic mass per proton, \sub{N}{H} is the hydrogen column density, $m_p$ is the proton mass, and $v$ is the outflow velocity. Table~\ref{tab:res} contains the calculated energetics that are similar for each outflow.

\begin{deluxetable*}{lcccccccc}
\tabletypesize{\footnotesize}
\tablecaption{Physical Properties, Distances, and Energetics of the four outflow systems\label{tab:res}}
\tablewidth{0pt}
\tablehead{
\colhead{Outflow System} & \multicolumn{2}{c}{S1 = --4900~km~s$^{-1}$} & \multicolumn{2}{c}{S2 = --5150~km~s$^{-1}$} & \multicolumn{2}{c}{S3 = --5350~km~s$^{-1}$} & \multicolumn{2}{c}{S4 = --5650~km~s$^{-1}$}\\
\colhead{Ionization Phase} & \colhead{Very High} & \colhead{High} & \colhead{Very High} & \colhead{High} & \colhead{Very High} & \colhead{High} & \colhead{Very High} & \colhead{High}
}
\startdata
log$_{}$(\sub{N}{H}) & 21.08\pme{0.41}{0.60} & 20.28\pme{0.40}{0.23} & 21.50\pme{0.17}{0.16} & 19.93\pme{0.24}{0.23} & 21.46\pme{0.25}{0.24} & 20.57\pme{0.47}{0.53} & 21.28\pme{0.32}{0.26} & 19.79\pme{0.21}{0.22}\\
(cm$^{-2}$)\\
\tableline
log$_{}$(\sub{U}{H}) & 0.3\pme{0.5}{0.1} & -0.8\pme{0.3}{0.1} & 0.6\pme{0.1}{0.1} & -0.7\pme{0.3}{0.2}  & 0.6\pme{0.1}{0.1} & -0.3\pme{0.2}{0.6} & 0.6\pme{0.1}{0.1} & -0.8\pme{0.3}{0.1} \\
(dex)\\
\tableline
log(\sub{n}{e}) & \tablenotemark{a}2.8\pme{0.3}{0.6} & 3.9\pme{0.1}{0.3} & \tablenotemark{a}2.7\pme{0.3}{0.3} & 4.0\pme{0.1}{0.2} & \tablenotemark{a}3.2\pme{0.3}{0.6} & 4.2\pme{0.1}{0.4} & \tablenotemark{a}2.5\pme{0.4}{0.5} & 3.9\pme{0.1}{0.2} \\
(cm$^{-3}$)\\
\tableline
Distance & \multicolumn{2}{c}{460\pme{200}{130}} & \multicolumn{2}{c}{360\pme{130}{100}} & \multicolumn{2}{c}{180\pme{220}{50}} & \multicolumn{2}{c}{460\pme{160}{140}} \\
(pc)\\
\tableline
$\dot{M}$ & \multicolumn{2}{c}{180\pme{310}{120}} & \multicolumn{2}{c}{350\pme{260}{170}} & \multicolumn{2}{c}{180\pme{320}{90}} & \multicolumn{2}{c}{300\pme{380}{170}} \\
($M{\astrosun}$yr$^{-1}$)\\
\tableline
log($\dot{\sub{E}{K}}$)\tablenotemark{b} & \multicolumn{2}{c}{45.14\pme{0.43}{0.51}} & \multicolumn{2}{c}{45.46\pme{0.25}{0.28}} & \multicolumn{2}{c}{45.21\pme{0.45}{0.31}} & \multicolumn{2}{c}{45.47\pme{0.36}{0.35}}\\
(erg~s$^{-1}$)\\
\tableline
$\dot{\sub{E}{K}}/L_{edd}$ & \multicolumn{2}{c}{1.1\pme{2.4}{0.8}} & \multicolumn{2}{c}{2.3\pme{3.1}{1.4}} & \multicolumn{2}{c}{1.3\pme{3.0}{0.8}} & \multicolumn{2}{c}{2.3\pme{4.4}{1.5}} \\
(\%)\\
\tableline
log(\ssub{f}{V}) & \multicolumn{2}{c}{-1.9\pme{0.79}{0.69}} & \multicolumn{2}{c}{-2.9\pme{0.43}{0.36}} & \multicolumn{2}{c}{-1.8\pme{0.57}{0.84}} & \multicolumn{2}{c}{-2.9\pme{0.46}{0.41}} \\
\enddata
\tablenotetext{a}{Assuming that both ionization components are at the same distance.}
\tablenotetext{b}{Assuming $\Omega$ = 0.4 and where $N_H$ is the sum of the two ionization phases.}
\tablecomments{Bolometric luminosity of \sub{L}{bol} = 1.3$^{+0.1}_{-0.1}\times10^{47}$~erg~s$^{-1}$ assuming the HE0238 SED.}
\end{deluxetable*}
\section{Discussion}
\label{sec:ds}
\subsection{Contribution to AGN Feedback}
To assess the potential for AGN feedback, \citet[]{hop10} and \citet[]{sca04} require kinetic luminosities exceeding 0.5\% or 5\% of the Eddington luminosity, respectively. Using the \sion{Mg}{ii}--based black hole mass equation from \citet[][]{bah19} and their methodology to measure the \sion{Mg}{II} FWHM and nearby continuum level from Sloan Digital Sky Survey (SDSS) data, the mass of the super massive black hole is $1.0^{+0.9}_{-0.5}\times10^{9} M_{\astrosun}$ (including systematics) with the corresponding Eddington luminosity (\sub{L}{edd}) of $1.3^{+1.1}_{-0.6}\times10^{47}$~erg~s$^{-1}$. Each outflow has a kinetic luminosity exceeding the requirement of \citet[]{hop10} with values above 1.0\% of \sub{L}{edd}. Collectively, the outflows have a kinetic luminosity of $8.8\times10^{45}$~erg~s$^{-1}$, which is 7.0\pme{6.5}{2.3}\% of \sub{L}{edd}. Therefore, these outflows, on average, carry enough energy to contribute substantially to AGN feedback in galaxies with similar black hole masses as in quasar 2MASS J1051+1247. For example, recent theoretical modeling has shown that BAL outflows effectively quench star formation within the host galaxy \cite[e.g.,][effective radii up to 10 kpc]{cho18} and drive gas into the intergalactic medium \cite[e.g.,][]{bre18}.
\subsection{Photoionization Solution and \sub{n}{e} Accuracy}
As seen in Figure~\ref{fig:sol}, the photoionization solutions for both phases in S1 and S4, as well as the high phases (HPs) in S2 and S3, are constrained only by \sub{N}{ion} upper and lower limits, which are immune to saturation effects. The multitude of upper and lower limits tightly constrain the errors in the solutions. The photoionization solutions of the HPs determined the upper limits (and sometimes values) of the resonance state \sub{N}{ion} for \sion{Ne}{v} and \sion{Ne}{vi} used to calculate the population ratios that yielded \sub{n}{e} for each outflow (see section~\ref{sec:ed}). When multiple diagnostics were available for a given outflow, the \sub{n}{e} values were all consistent within errors. This consistency in \sub{n}{e} between multiple diagnostics along with the tightly constrained photoionization solutions shows the results are robust and accurate (for more discussions on these issues, see Paper I).   
\subsection{Geometry and Volume Filling Factor}
There are striking similarities in the geometry between outflows. We assume the VHP occupies the space where the outflow resides \cite[][]{ara13}. Therefore, the VHPs of S1 and S3 have the same thickness (\sub{N}{H}/\sub{n}{e}) of 0.62 pc. Similarly, S2 and S4 have a thickness of 2.0 pc for their VHP. All outflows have thicknesses less than 0.6\% of their calculated distances. The similarities continue when looking at the volume filling factor of the HP in each outflow (the VHP volume filling factor = 1 given our assumption). Due to the kinematic similarities of the troughs from the VHP and HP for each outflow, it is physically plausible that the two phases are occupying the same volume. Since the HP is both denser and has a lower \sub{N}{H} than that of the VHP, the HP has to have a small volume filling factor within the VHP. This volume filling factor is given by equation 6 in Paper II (see also Section 2.5 in Paper I):   
\begin{equation}
\ssub{f}{V} = \frac{\sub{U}{H,\tiny\textit{HP}}}{\sub{U}{H,\tiny\textit{VHP}}}\times\frac{\sub{N}{H,\tiny\textit{HP}}}{\sub{N}{H,\tiny\textit{VHP}}}
\end{equation}
S1 and S3 have \ssub{f}{V} values differing by 25\% without considering errors, and S2 and S4 have the same \ssub{f}{V} value (see Table~\ref{tab:res}). Considering the errors, all four outflows have consistent \ssub{f}{V} values. These similarities in geometry, the consistent distances, and small velocity separations suggest the outflows have a common origin. 
\subsection{Connection to X-Ray Warm Absorbers}
The two-phase solutions required for each outflow to sufficiently reproduce the observed \sub{N}{ion} from the high- and very high-ionization potential ions are similar to what is seen for X-ray warm absorbers. The ionization parameter of X-ray warm absorbers can span up to five orders of magnitude (-1 $<$ log($\xi$) $<$4) and necessitate a continuous hydrogen column density as a function of $\xi$ \cite[e.g.,][]{ste03,cos07,hol07,mck07,beh09}. The current data allows for higher-ionization phases to exist within the outflows. For the HE0238 SED, log($\xi$) $\approxeq$ log(\sub{U}{H})+1.3, and, therefore, the photoionization solutions are comparable to what is determined for X-ray warm absorbers.
\subsection{The ``Shading Effect"}
\label{sec:tp}
Even though the four outflow systems may reside at similar distances, the SED seen by an exterior outflow will be attenuated by an interior one \cite[e.g.,][]{bau10,sun17,mil18}. To test the effects this may have on the results, we used the same approach as \cite{mil18} and assumed that S3 shades the other outflows since it has the smallest calculated distance and the largest total column density. New Cloudy model grids were generated using the transmitted SEDs from both the high and very high photoionization solutions for S3. New photoionization solutions, distances, and energetics were determined for the other outflows (see elaboration in Paper V). The end result was that the distance and energetics decreased by less than 15\%, which is small compared to the overall errors.
\section{Summary and Conclusions}
\label{sec:sc}
This paper presented new \textit{HST}/COS spectra for the quasar 2MASS J1051+1247, which contains four outflow systems. For the first time, we identified absorption troughs from transitions of \sion{Ar}{vi} 544.73~\AA\ and 548.90~\AA, \sion{O}{IV*} 555.26~\AA, \sion{Ne}{vi*} 562.80~\AA, and \sion{Ne}{v*} 569.80~\AA\ and 572.30~\AA. The absorption troughs yielded ionic column density measurements/lower limits for up to 11 ions in each outflow system. Best-fit photoionization solutions (\sub{U}{H} and \sub{N}{H}) were determined for each outflow by using a grid of photoionization models in conjunction with the ionic column density constraints. 

Column density ratios between two excited states and the ground state of \sion{Ne}{v} as well as one excited state and ground state of \sion{Ne}{vi} yielded consistent \sub{n}{e} for the outflows with multiple determinations (see Figure~\ref{fig:dens}). These electron number densities were used in equation (\ref{eq:R}) to calculate the distance to the central source of each outflow. The mass flux and kinetic luminosity of each outflow were determined from the distance and equations (\ref{eq:M})~\&~(\ref{eq:E}). Finally, AGN feedback was assessed, and all of these results are shown in Table~\ref{tab:res}.

The following emerges from this work:
\begin{enumerate}
\item The never-before-seen ionic transitions from \sion{Ar}{vi}, \sion{O}{iv*}, \sion{Ne}{vi*}, and \sion{Ne}{v*} were revealed by the EUV500 \textit{HST}/COS observations. The \sion{Ne}{vi*} and \sion{Ne}{v*} identifications enabled the electron number density, distance, and energetics of all outflows to be determined.
\item A two-phase ionization solution is needed in each outflow to simultaneously satisfy the column density measurements from ions with a wide range of ionization potentials (80--520 eV).
\item The small velocity separations, consistent distances within the errors, and other geometric similarities suggest the outflows originate from the same material at the same distance.
\item The outflows individually (depending on the theoretical work) and collectively have a large enough kinetic luminosity to Eddington luminosity ratio to be major contributors to AGN feedback processes.
\end{enumerate}


\acknowledgments

T.M., N.A., and X.X. acknowledge support from NASA  grants \textit{HST} GO 14777, 14242, 14054, and 14176. This support is provided by NASA through a grant from the Space Telescope Science Institute, which is operated by the Association of Universities for Research in Astronomy, Incorporated, under NASA contract NAS5-26555. T.M. and N.A. also acknowledge support from NASA ADAP 48020 and NSF grant AST 1413319. CHIANTI is a collaborative project involving George Mason University (USA), the University of Michigan (USA), and the University of Cambridge (UK).



\end{document}